\documentclass[aps,amsfonts]{revtex4}
\usepackage{epsfig}
\usepackage{graphicx}
\usepackage{amsmath}
\usepackage{amsbsy}
\begin{document}
\newcommand{\ee}{\end{equation}}
\newcommand{\be}{\begin{equation}}
\newcommand{\bea}{\begin{eqnarray}}
\newcommand{\eea}{\end{eqnarray}}
\def\x{\mathbf{x}}
\def\p{\mathbf{p}}
\def\ho{{\mbox{\tiny{HO}}}}
\def\sc{\scriptscriptstyle}
\title{Non-commutative Supersymmetric Quantum Mechanics}
\author{  Ashok  Das$^{a,b}$, H.  Falomir$^{c}$,  J. Gamboa$^{d}$  and
  F.           M\'endez$^{d}$}           \email{das@pas.rochester.edu,
  falomir@fisica.unlp.edu.ar, jgamboa55@gmail.com, feritox@gmail.com}
\affiliation{$^{a}$Departament of Physics and Astronomy, University of
  Rochester, USA}
\affiliation{$^{b}$Saha    Institute   of   Nuclear    Physics,   1/AF
  Bidhannagar, Calcutta-700 064, India}
\affiliation{$^{c}$IFLP/CONICET   -   Departamento   de   F\'{\i}sica,
  Facultad de Ciencias Exactas,
Universidad Nacional de la Plata, C.C. 67, (1900) La Plata, Argentina}
\affiliation{$^{d}$Departamento   de   F\'{\i}sica,   Universidad   de
  Santiago de Chile, Casilla 307, Santiago, Chile}

\begin{abstract}
General non-commutative supersymmetric quantum mechanics models in two
and three dimensions are constructed and some two and three dimensional examples are explicitly studied. The structure of
the theory studied
suggest other possible applications in physical systems with potentials involving spin and non-local interactions.

\end{abstract}
\maketitle

\section{Introduction} \label{Introduction}
The non-commutativity  of spacetime is  an old idea \cite{old}  and in
physics, the first example  of noncommutativity was probably discussed
by Landau in  1930 \cite{landau}. It has been  revived in recent years
within  the  context of  string  theory  \cite{alot}  and since  then,
non-commutative field  theories have attracted  much  attention in
various fields  such as mathematical  physics \cite{connes,review} and
phenomenology  \cite{hinf}. Normally,  in discussing  a  quantum field
theory  on a  non-commutative manifold,  we assume  that  only spatial
coordinates do  not commute in  order to avoid possible  conflict with
unitarity of  the theory. On  a manifold where spatial  coordinates do
not commute, the product of functions generalizes to the Moyal product
(also  conventionally  known  as  the  $\star$-product)  which  is  an
associative and non-commutative product defined as
\begin{equation}
\label{moyal}
(A\star  B)(\x)  =  e^{ \frac{i}{2}  \,  \theta^{ij}  \,
  \partial^{(1)}_i \,
    \partial^{(2)}_j } A(\x_1)B(\x_2)\bigg{|}_{\x_1=\x_2=\x},
\end{equation}
where  $A,B$  denote two  arbitrary  functions  of  coordinates and  $
\theta^{ij} =- \theta^{ji}$ denotes the constant parameter of non-commutativity
with   $i,j=1,2,\cdots,n$  taking   values  over   only   the  spatial
indices. It follows from \eqref{moyal} that
\begin{equation}
[x^{i},x^{j}]_{\star} := x^{i}\star x^{j} - x^{j}\star
x^{i} = i \theta^{ij},\label{basicnc}
\end{equation}
which reflects the basic spatial non-commutativity of the manifold.

A  particularly  interesting example  of  non-commutative theories  is
non-commutative  quantum  mechanics   (NCQM)  \cite{ncqm}  which contains
most of  the distinguishing  properties of  a non-commutative
quantum  field  theory   and  yet  is  simple  enough   to  carry  out
explicit calculations.   At   the   quantum   mechanical   level,   the   space
non-commutativity   \eqref{basicnc}  reflects  in   the  Schr\"odinger
equation with  the change  of the usual  products of functions  by the
$\star$-product    \eqref{moyal}.   Namely,    the    time   dependent
Schr\"{o}dinger equation in a non-commutative space takes the form
\be
i\frac{\partial  \phi (\mathbf{x},t)}{\partial t}=  \left[ -\frac{1}{2}
  \mbox{\boldmath$\nabla$}^2 + V(\mathbf{x}) \right] \star \phi (\mathbf{x},t)
= -\frac{1}{2} \mbox{\boldmath$\nabla$}^2 \phi (\mathbf{x},t) + V(\mathbf{x}) \star \phi
(\mathbf{x},t)\,,\label{tdseqn}
\ee
where  $\phi  (\mathbf{x},t)$  denotes  the  quantum  mechanical  wave
function (and we have set $\hbar = 1= m$). We note from \eqref{moyal} that the Moyal product definition
does not  affect the  derivatives and the  only change appears  in the
potential term which can be simplified using \eqref{moyal} to have the
form (we are suppressing the  time dependence of the wave function for
simplicity)
 \be \label{Bopp}
    V(\mathbf{x}) \star \phi (\mathbf{x}) =
    \frac{1}{(2 \pi)^{n/2}} \int d^n k \, \widetilde{V}(\mathbf{k})\,
    e^{\imath {k_i} \big[{x^i} -  \frac{\theta^{i j}\, p_j }{2} \big]} \,
    {\phi}(\mathbf{x})
    =:    V\left(\mathbf{x}   -{\overline{\mathbf{p}}}/{2}\right)   \,
    \phi(\mathbf{\mathbf{x}})\,,
\ee
where  $\widetilde{V} (\mathbf{k})$ denotes  the Fourier  transform of
the potential,  $p_i:= - \imath \partial/\partial  x^i$ represents the
momentum operator and we  have identified $\overline{p}^i := \theta^{i
  j} p_j$  for simplicity. The simple  replacement of $x^i$  by $x^i -
\overline{p}^i/{2}$  in \eqref{Bopp} in  going from  a non-commutative
star product to an ordinary product is known as the Bopp shift. Notice
that there is no ordering ambiguity in this definition since
\begin{equation}\label{a5}
    \left[    \mathbf{k}\cdot    \mathbf{x}    ,   \mathbf{k}    \cdot
      \overline{\mathbf{p}} \right]
    = k_i k_j \theta^{j k} \left[ {x^i} , {{p_k}} \right]
    = k_i k_j \theta^{j k} \, \imath\,  \delta^{i}_{ k}
    =- \imath\,  k_i k_j \theta^{i j} =0\,,
\end{equation}
where we have used $[p_{i},x^{j}] = -\imath (\partial_{i} x^{j}) =
- \imath \delta^{j}_{i}$. The procedure  sketched here was employed in
\cite{glr}  to illustrate  some of  the properties  of non-commutative
quantum  field  theories using  a  general  realization  of NCQM.   In
particular, several  interesting properties of  the central potentials
within the context of NCQM were derived using this formalism.

Another class  of quantum  mechanical theories that  have been  studied from
various   points  of   view   is  known   as  supersymmetric   quantum
mechanics. In  this connection  much work has  been done  within the context of  one
dimensional  supersymmetric  quantum  mechanical  theories  with  some
isolated  attempts  at  supersymmetric  quantum  mechanics  in  higher
dimensions.  The goal  of the  present  work is  to construct  general
(higher dimensional) non-commutative supersymmetric quantum mechanical
systems  and, in particular,  to study  three-dimensional  non-commutative
problems.  Although the  supersymmetric extensions  of non-commutative
quantum  mechanical   systems  have  been  partly   discussed  in  the
literature  (see, for  example, \cite{hari}  and \cite{ghosh}  for the
harmonic oscillator  and the Landau problems), a  general formalism to
study  such systems  is  still lacking,  and  we wish  to systematically
develop it in this work.

Our  paper is  organized as  follows. In  section {\bf  II} we  give a
general  discussion of Witten's  supersymmetric quantum  mechanics (in
ordinary space) and discuss in detail several subtle points related to
its extension  to higher (three) dimensional systems.  In section {\bf
  III}  we generalize this to non-commutative space and construct general supersymmetric theories in higher dimensional non-
  commutative spaces. We study, as examples,  the
non-commutative     supersymmetric     harmonic    oscillator,     the
non-commutative supersymmetric  Landau problem  as well as the  three  dimensional harmonic oscillator.  We present
our conclusions and discussions in section {\bf IV} and collect some technical results in two appendices.

\section{Supersymmetric Quantum Mechanics} \label{3-SUSYQM}
One dimensional  supersymmetric quantum  mechanics  (SUSYQM) has  been
studied extensively in  the past couple of decades  (for a review, see
\cite{khare}).  While  in  the  beginning  it  drew  attention  as  an
interesting toy  model for studying the  mechanism of supersymmetry breaking,
since  then it  has found  many interesting  applications.  Basically,
SUSYQM in one dimension is described by the simple closed superalgebra
(graded algebra with curly brackets denoting anti-commutators)
\begin{equation}\label{su1}
    \begin{array}{c}
      \left\{Q,Q^\dagger\right\}= 2 H \,,
      \\ \\
      \left[Q,H\right]  =0\,, \quad \left[Q^\dagger,H\right]= 0\,,
      \\ \\
      \left\{Q, Q \right\}=0\,, \quad \left\{Q^\dagger,Q^\dagger\right\}=0 \, ,
    \end{array}
\end{equation}
where  the  supercharges $Q,  Q^{\dagger}$  define  the Grassmann  odd
generators of the algebra.
In one dimension the supercharges can be given the coordinate representation
\begin{equation}\label{supercharges}
    Q:= A \psi = \left( \frac{d}{dx} +W(x) \right) \psi \,, \quad
    Q^\dagger  :=  A^{\dagger}  \psi^{\dagger}  =  \left(-\frac{d}{dx}
    +W(x) \right) \psi^\dagger\,,
\end{equation}
where $W(x)$ is chosen to be real and is known as the  superpotential and  $\psi$  and
$\psi^\dagger$ are nilpotent Grassmann odd elements satisfying the algebra
\be
\{\psi,\psi^\dagger\}  =1\, , \quad \{\psi, \psi\}=0=\{\psi^\dagger,
\psi^\dagger\}. \label{su4}
\ee
In  one dimension  the algebra  (\ref{su4})  can  be  explicitly
realized through the identification
\be
\psi   \rightarrow  \sigma_+\, ,  \quad  \psi^\dagger  \rightarrow
\sigma_-,
\label{su5}
\ee
where  $\sigma_\pm =  (\sigma_1 \pm  i \sigma_2)/2$  with $\sigma_k\,,
k=1,2,3,$ denoting the  Pauli matrices. The supersymmetric Hamiltonian
can now be constructed from \eqref{su1} to have the form
\begin{equation}
2   H   =   -   \frac{d^{2}}{dx^{2}}   +  W^{2}   (x)   +   W^{\prime}
(x)\,\sigma_{3},\label{1dsusyH}
\end{equation}
with the supersymmetric partner Hamiltonians of the form
\begin{equation}
2H_{\pm} = - \frac{d^{2}}{dx^{2}} + W^{2} (x) \pm W^{\prime} (x),
\end{equation}
where  ``prime" denotes  differentiation with  respect to  $x$.  It is
worth emphasizing here that the  nilpotency of the supercharges in the
explicit  realization   \eqref{supercharges}  is  guaranteed   by  the
nilpotency $\psi^{2} = 0  = (\psi^{\dagger})^{2}$ of the Grassmann odd
elements.

We note  here that although our  discussion here has  assumed that the
superpotential  is a  real function  as is  conventional, this  can be
relaxed. If $W(x)$ is a complex superpotential, then we can write
\begin{equation}
Q =  \left(\frac{d}{dx} + W (x)\right)  \sigma_{+},\quad Q^{\dagger} =
\left(-\frac{d}{dx}            +            W^{*}           (x)\right)
\sigma_{-},\label{1dcomplexsuperpotential}
\end{equation}
which leads to the supersymmetric Hamiltonian
\begin{eqnarray}
2H = \left\{Q,  Q^{\dagger}\right\} & = & -  \frac{d^{2}}{dx^{2}} - 2i
{\rm Im}\, W (x)\,\frac{d}{dx} -  i {\rm Im}\, W^{\prime} (x) + WW^{*}
+ {\rm Re}\,
W^{\prime} (x) \sigma_{3}\nonumber\\
&     =    &     -\left(\frac{d}{dx}    +     i    {\rm     Im}\,    W
(x)\right)^2 + \left({\rm
  Re}\,    W   (x)\right)^{2}    +   {\rm    Re}\,    W^{\prime}   (x)
\sigma_{3},\label{1dcomplexH}
\end{eqnarray}
with the supersymmetric partner Hamiltonians of the form
\begin{equation}
2H_{\pm} = - \left(\frac{d}{dx} + i {\rm Im}\, W (x)\right)^{2} + ({\rm Re}\, W(x))^{2} \pm {\rm Re}\, W^{\prime} (x).
\end{equation}
The  Hamiltonian   is  by  construction  Hermitian   even  though  the
superpotential  is  complex  and  \eqref{1dcomplexH}  shows  that  the
imaginary  part of the  superpotential behaves  like a  gauge potential,
${\cal  A} =  -{\rm Im}\,  W$, and  when the  superpotential  is real,
\eqref{1dcomplexH}   reduces  to  \eqref{1dsusyH}.   Furthermore,  the
zero-energy ground state of the theory, satisfying $Q\phi_{0} (x)=0$ or $Q^\dag\phi_{0} (x)=0$ with the nonzero element
proportional to (note that $\phi_{0}$ is a two component wave function)
\begin{equation}
\Phi_{0} (x) \sim
e^{\int^{x} dx'\left(\mp{\rm Re}\, W(x') + i {\rm Im}\, W(x')\right)},
\end{equation}
is normalizable only if  ${\rm Re}\, W(x)$ is an odd function
of $x$ which does not vanish too fast for $|x|\rightarrow \infty$,
without any restriction  on the  imaginary part of  the
superpotential. Simple examples of
unbroken supersymmetry are given by the  superpotentials $W (x) = x + i
a$, $W(x) = x^{3} + i a x^{2}$ and so on.

\medskip

This  construction   can  be  generalized  to   two  dimensions  quite
easily.  Let us label   the  two  dimensional  coordinates  as  $x^{k},   k=1,2$ and
introduce  two complex  superpotentials  $W_{k}$. Defining $A_k:= \partial_k
+ W_k(x)$ and identifying $\psi_{k} = (-1)^{k+1}\,\frac{\sigma_{k}}{2}$, we  can write  the
manifestly rotation invariant supercharges as ($k,l=1,2$ and repeated
indices are summed)
\begin{eqnarray}
Q   &=&   A_{k}\left( \delta_{k l}  -   i   \epsilon_{kl} \right) \psi_{l}   =
\left(\partial_{k}    +   W_{k}\right) \left( \delta_{k l}  -   i   \epsilon_{kl} \right) (-1)^{l+1} \frac{\sigma_{l}}{2}    \nonumber\\
&=&  \left(A_1+ \imath A_2\right)  \sigma_+  =  \left(\partial_1  +  i  \partial_2  +
\left(W_{1}+iW_{2}\right)\right) \sigma_{+},\nonumber
\\
Q^{\dagger} &=  & \left(A_1^{\dagger}- \imath A_2^{\dagger}\right) \sigma_- = \left(-\partial_1
 + i \partial_2 + (W_{1}^{*}-iW_{2}^{*})\right)\sigma_{-},
\label{2dcomplexQ}
\end{eqnarray}
where $\epsilon_{12}=1=-\epsilon_{21}$ and $\epsilon_{11}=0=\epsilon_{22}$.

Notice that the   two   complex  supercharges   are nilpotent by construction, and they can also be written in the form
\begin{equation}\label{Q-gauge}
    \begin{array}{c}
      Q= \left( \left(\partial_1 - \imath \mathcal{A}_1 \right) +
      \imath \left( \partial_2 - \imath \mathcal{A}_2 \right) \right) \sigma_+\,,
       \\ \\
      Q^\dag= \left( - \left(\partial_1 - \imath \mathcal{A}_1 \right) +
      \imath \left( \partial_2 - \imath \mathcal{A}_2 \right)  \right) \sigma_-\,,
    \end{array}
\end{equation}
where we have introduced the equivalent gauge fields
\begin{equation}
{\cal  A}_{k}  =  -({\rm  Im}\,  W_{k} +  \epsilon_{kl}\,  {\rm  Re}\,
W_{l}).\label{h2-A}
\end{equation}

The supersymmetric Hamiltonian can now be easily obtained from the
superalgebra
\begin{eqnarray}
2 H  = \left\{Q, Q^{\dagger}\right\} &= -   \left(\partial_1  -  i  {\cal  A}_{1}\right)^{2}  -
\left(\partial_2  - i  {\cal A}_{2}\right)^{2}  +  {\cal F}_{12}
\sigma_{3}   =   -   \mathbf{D}\cdot   \mathbf{D}  +   {\cal   F}_{12}
\sigma_{3},
\label{2dcomplexH}
\end{eqnarray}
where we have identified the  field strength
\begin{equation}
{\cal F}_{12} =  \partial_{1} {\cal A}_{2} - \partial_{2}
{\cal A}_{1}.\label{h2}
\end{equation}
Therefore,  this  supersymmetric system  behaves  like  a minimally  gauge
coupled system  with a magnetic dipole interaction.  The planar Landau
model is a typical example of such a system. Notice that only the
combinations of real and imaginary parts of the superpotentials in
Eq.\ (\ref{h2-A})
enter in the expression of the Hamiltonian (as well as the supercharges).

\medskip

Although  more  realistic physical  systems  with supersymmetry  would
correspond to  three dimensional quantum mechanical  systems
\cite{iachello,nuclear, alemanes}, surprisingly there is only a
few isolated discussions about
higher dimensional  SUSYQM. A construction similar  to the discussions
above can be  generalized to three dimensions in  ordinary space using
suitable generalizations of (\ref{su4}) and (\ref{su5}). In this case,
the space  of the Grassmann elements  needs to be  enlarged. Indeed, in
three dimensions  we can proceed  as in \cite{gz}.  Introducing  three
$A_{k}$'s  with
$k=1,2,3$  corresponding to the  three coordinates,  we can  write the
supercharges as
\begin{eqnarray}
\label{susy3}
Q  = \sum_{k=1}^3 A_k \psi_k\, ,\quad
Q^{\dagger} = \sum_{k=1}^3 A^{\dagger}_k \psi^{\dagger}_k\,,
\end{eqnarray}
where   as  in   \eqref{supercharges}   we  have   assumed  that   the
superpotential is complex and have identified
\begin{equation}\label{SuperQ}
    A_k:= \left(  \partial_k  + W_k(\mathbf{x}) \right) \,, \quad
    A^{\dagger}_k:= \left( -\partial_k  + W_k^{*} (\mathbf{x}) \right)
    \,, \quad k=1,2,3\,.
\end{equation}
Clearly, in this case,  we have three distinct complex superpotentials
$W_{k}  (\mathbf{x})$  and   six  Grassmann  odd  elements  $\psi_{k},
\psi^{\dagger}_{k}$.

The nilpotency of the supercharges in \eqref{susy3} requires that
\begin{equation}
\psi_{k}\psi_{l} = 0 = \psi^{\dagger}_{k}\psi^{\dagger}_{l},
\label{psinilpotency}
\end{equation}
for  any pair  of $k,l  = 1,2,3$  which in  turn implies  the standard
Grassmann property
\begin{equation}
\{\psi_{k},\psi_{l}\}                =               0               =
\{\psi^{\dagger}_{k},\psi^{\dagger}_{l}\}.\label{Grassmann}
\end{equation}
However, it  is worth noting that \eqref{psinilpotency}  is a stronger
condition  than \eqref{Grassmann}  and  this seems  not  to have  been
appreciated in the  past. A realization of the  Grassmann odd elements
satisfying \eqref{psinilpotency} leads to
\be
\label{real}
\psi_k =  \sigma_k \otimes \sigma_+\, ,  \quad \psi^\dagger_k= \sigma_k
\otimes \sigma_-\, , \quad k=1,2,3\,,
\ee
and the  supersymmetric Hamiltonian $H$  can now be obtained  from the
algebra \eqref{su1} to correspond to
\begin{equation}\label{H-SUSY}
    2 H:=  \left\{ Q , Q^\dagger \right\}
    = \frac{1}{2}\left\{  A_k , A_l^\dagger \right\}  \left\{ \psi_k ,
    \psi_l^\dagger \right\}
    +  \frac{1}{2}\left[ A_k  ,  A_l^\dagger \right]  \left[ \psi_k  ,
      \psi_l^\dagger \right]\,.
\end{equation}
Using  the   explicit  realization   of  the  Grassmann   elements  in
\eqref{real} we now obtain
 \bea
\{ {\psi_k}  ,  \psi_l^\dagger \}=  \delta_{k  l} \  {\mathbf{1}}_2
\otimes  \mathbf{1}_2 +\imath \, \epsilon_{k  l m  }\, \sigma_m  \otimes
\sigma_3\, ,
\nonumber
\\
\left[ {\psi_k}  ,  \psi_l^\dagger \right]=  \delta_{k  l} \  {\mathbf{1}}_2
\otimes  \sigma_3 +\imath  \,  \epsilon_{k l  m  }\, \sigma_m  \otimes
\mathbf{1}_2\, , \label{conmut-psi}
\eea
and   substituting  these   into  (\ref{H-SUSY})   the  supersymmetric
Hamiltonian for the system takes the general form
\begin{eqnarray}
2 H &=& \left(- (\partial_{k} +  i\, {\rm Im} W_{k})^{2} + ({\rm Re}\,
W_{k})^{2}\right)  \mathbf{1}_2 \otimes  \mathbf{1}_2  + (\partial_{k}
{\rm Re}\, W_{k}) \mathbf{1}_{2} \otimes \sigma_{3}\nonumber
\\
 & & + \epsilon_{klm} \left(i \Big((\partial_{k} {\rm Re}\, W_{l}) + 2
  {\rm  Re}\, W_{l}  \partial_{k} -  2i\, {\rm  Im}  W_{k}\,{\rm Re}\,
  W_{l}  \Big)\sigma_m  \otimes\sigma_3  +  (\partial_{k}  {\rm  Im}\,
  W_{l}) \sigma_{m}\otimes \mathbf{1}_{2}\right)\nonumber
\\
 &  =  & \left(-  \mathbf{D}^{2}  +  ({\rm Re}  \mathbf{W})^{2}\right)
\mathbf{1}_{2}\otimes  \mathbf{1}_{2} + (\mbox{\boldmath$\nabla$}\cdot
       {\rm Re} \mathbf{W}) \mathbf{1}_{2} \otimes \sigma_{3}\nonumber
\\
 & & -i\left({\rm  Re} \mathbf{W}\times \mathbf{D} - \mathbf{D}
\times   {\rm  Re}  \mathbf{W}\right) \cdot  \boldsymbol{\sigma}  \otimes
\sigma_{3}-\frac{1}{2}  \epsilon_{klm}  {\cal F}_{kl}\sigma_m  \otimes
\mathbf{1}_{2}
\label{H-SUSY1}
\end{eqnarray}
where we have identified
\begin{equation}
\mathbf{D}  =  \mbox{\boldmath$\nabla$}  +  i {\rm  Im}  \mathbf{W}  =
\mbox{\boldmath$\nabla$}  - i  \mbox{\boldmath${\cal  A}$},\quad {\cal
  F}_{ij} = \partial_{i} {\cal A}_{j} - \partial_{j} {\cal A}_{i}.
\end{equation}
The  supersymmetric  partner Hamiltonians  can  now  be obtained  from
\eqref{H-SUSY1} to correspond to
\begin{eqnarray}
2H_{\pm} & = & \left(-  \mathbf{D}^{2} + ({\rm Re} \mathbf{W})^{2} \pm
(\mbox{\boldmath$\nabla$}\cdot     {\rm     Re}     \mathbf{W})\right)
\mathbf{1}_{2}\nonumber
\\
& &- \frac{1}{2}\,\epsilon_{klm} {\cal F}_{kl}\sigma_m \mp i \left(
{\rm  Re}  \mathbf{W}\times  \mathbf{D}   -  \mathbf{D}  \times  {\rm  Re}
\mathbf{W}\right) \cdot {\boldsymbol \sigma}.
\end{eqnarray}

We note  here that  a sufficient condition  for the invariance  of the
Hamiltonian under  supersymmetry transformations is  the nilpotency of
the supercharges,  which is ensured  by \eqref{psinilpotency}. Indeed,
from    (\ref{H-SUSY})   it    follows   that    under   supersymmetry
transformations   with    Grassmann   parameters   $\varepsilon$   and
$\overline{\varepsilon}$, the change in the Hamiltonian is given by
\begin{equation}\label{H-inv-nilp}
    \Big[ {\varepsilon} Q + \overline{\varepsilon} Q^\dagger, H \Big] =
     {\varepsilon} \Big[ Q^2, Q^\dagger \Big] +
     \overline{\varepsilon}  \left[ \left(Q^\dagger\right)^2, Q \right] \,,
\end{equation}
which clearly  vanishes if  $Q^{2} = 0  = (Q^{\dagger})^{2}$.  We will
show this in detail in appendix \ref{SUSYinvariance} and simply note here that this
continues to be true in the non-commutative case as well.

\section{Non-commutative supersymmetric quantum mechanics} \label{NC-3-SUSYQM}

In this section  we will extend the construction  of general SUSYQM of
the  last   section  to  non-commutative  space.  We   will  take  the
supersymmetry  algebra   to  correspond  to  the   graded  algebra  of
\eqref{su1} except that the brackets would now involve $\star$-product
defined in \eqref{moyal}. Thus, for example, we have
\begin{equation}\label{H-SUSY-NC}
    2   H^{\sc (NC)}  :=  \{Q,Q^\dag\}_\star=Q\star   Q^\dag+Q^\dag\star  Q,\quad
    \{Q,Q\}_{\star} = 0 = \{Q^{\dagger},Q^{\dagger}\}_{\star}.
\end{equation}
We note  that the non-commutative  spaces we are interested in are of spatial
dimension two  or higher  (since we are  considering non-commutativity
only for space coordinates).

\subsection{Two dimensional case}

Let  us consider a  simple non-commutative  manifold with  two spatial
dimensions. In such a case,  the parameter of non-commutativity can be
characterized   by   a   single   parameter   $\theta$   through   the
identification
\begin{equation}
\theta^{ij} = \theta \epsilon^{ij},\quad i,j=1,2,
\end{equation}
so that the basic commutator of coordinates \eqref{basicnc} reduces to
\begin{equation}
[x^i,x^j]_\star = \imath\, \theta\, \epsilon^{ij}\,.
\end{equation}
In such a case, the Bopp shift \eqref{Bopp} takes the simple form
\begin{equation}
x^i \rightarrow x^i -\frac{\theta}{2}\, \epsilon^{ij} p_j,\label{Bopp1}
\end{equation}
where  the  momentum operator  can  be  identified  in the  coordinate
representation with $p_{i} = -i \partial_{i}$.

To construct  the general non-commutative SUSYQM  model in two  dimensions, we
take    the   supersymmetric    charges    already   constructed    in
\eqref{2dcomplexQ}   with  two   general  complex   superpotentials.  The
Hamiltonian now can be obtained in a straightforward manner to be
\begin{eqnarray}
2H^{\sc (NC)} &=& \left\{Q, Q^\dagger \right\}_\star\nonumber
\\
&=&\left(- \left(\partial_1  - i {\cal A}_{1}\right)^{2}_{\star}
- \left(\partial_2  -  i {\cal  A}_{2}\right)^{2}_{\star}\right)
\mathbf{1}_{2} + \left(\partial_{1}  {\cal A}_{2} - \partial_{2} {\cal
  A}_{1}  - i  \left[{\cal A}_{1},  {\cal A}_{2}\right]_{\star}\right)
\sigma_{3}\nonumber
\\
& = & - \left(\mathbf{D}\cdot \mathbf{D}\right)_{\star} \mathbf{1}_{2}
+ {\cal F}_{12} \sigma_{3} , \label{h2-NC}
\end{eqnarray}
with the supersymmetric partner Hamiltonians
\begin{equation}
2 H^{\sc (NC)}_{\pm} = - (\mathbf{D}\cdot \mathbf{D})_{\star} \pm {\cal F}_{12},
\end{equation}
where ${\cal A}_{k}, k=1,2$ are defined in \eqref{h2} and we have identified
\begin{equation}
{\cal F}_{12} = \partial_{1}  {\cal A}_{2} - \partial_{2} {\cal A}_{1}
- i \left[{\cal A}_{1}, {\cal A}_{2}\right]_{\star},\label{NCfieldstrength}
\end{equation}
which is the  generalization of the field strength  tensor associated with the
Abelian gauge  field in a  non-commutative theory. The  Moyal products
can be evaluated using the Bopp shift \eqref{Bopp1}
\be
W_k (\mathbf{x}) \rightarrow W_k \left(x^1 - \frac{\theta}{2} \, p_2 ,
x^2+\frac{\theta}{2} \, p_1\right)\,, \label{cr}
\ee
so that \eqref{h2-NC} gives the  most general (depending on the choice
of $W_{k} (\mathbf{x})$) two dimensional supersymmetric Hamiltonian in
the non-commutative space.

We note  here that one of  the key ingredients in  solving exactly the
spectrum of the  one dimensional problems in the  conventional case is
the  connection  between  the  ground  state  wave  function  and  the
superpotential.   In   the   conventional  two-dimensional   problems,
something similar  can also be  done if one finds  normalizable ground
state   solutions   satisfying   $Q  \phi_0  (x)=0$   or   $Q^\dagger
\phi_0 (x)=0$. In  the non-commutative case, however,  the ground state
equations  involve Moyal  products which  cannot be  easily integrated
since  (contrary to  the conventional  case) they  contain arbitrarily
many derivatives because of the Bopp shift.

\medskip

As  an example of the general two dimensional theory, let  us consider  the  SUSY Landau
problem in  the non-commutative  space. The Hamiltonian  describes the
motion of a charged particle on  a plane moving under the influence of
a constant magnetic field. Therefore, in the commutative space we can choose the vector potentials in the symmetric gauge,
\begin{equation}
{\cal A}_{1} = - \frac{B}{2}\,x^{2},\quad {\cal A}_{2} = \frac{B}{2}\, x^{1},\label{symmgauge}
\end{equation}
where $B$ represents  the constant magnetic field. (We assume $B>0$ for simplicity.) This can
be achieved  ((see \eqref{2dcomplexQ}, \eqref{2dcomplexH})) by choosing, for example,
${\rm Re}\,  W_{i} = 0, {\cal  A}_{i} = -{\rm
  Im}\, W_{i}$. In this case, we have
\begin{eqnarray}
D_{i} &  = &  \partial_{i} + \frac{i}{2}\,  B \epsilon_{ij} x^{j}  = i
(p_{i} + \frac{B}{2}\,\epsilon_{ij} x^{j}),\nonumber
\\
Q &  = & i  \left(p_{1} + ip_{2} -  \frac{iB}{2} (x^{1}+ix^{2})\right)
\sigma_{+},\nonumber
\\
Q^{\dagger}   &   =   &   -  i   \left(p_{1}-ip_{2}   +   \frac{iB}{2}
(x^{1}-ix^{2})\right) \sigma_{-}.\label{landauQ}
\end{eqnarray}
With  these, the  Hamiltonian for  the Landau  problem  in commutative
space can be obtained to be (see \eqref{2dcomplexH})
\begin{equation}
\label{hamlan}
2H=\left({\bf      p}^2+\left(\frac{B}{2}\right)^2{\bf     x}^2     -
{B}\,L\right) \mathbf{1}_2 + B\sigma_3\,,
\end{equation}
where $L$ denotes the third  component of the orbital angular momentum defined
to be $L = (x^{1} p_{2} - x^{2} p_{1})$.

Let us  note \cite{glr2,we3} that the operator  within the parenthesis
in \eqref{hamlan}
\be
{\cal H} :=  {\bf p}^2 + \left(\frac{B}{2}\right)^{2} {\bf x}^2 - B \, L
= H_0 - B \, L \,, \label{w1}
\ee
can be  formally identified with the Hamiltonian  of a two-dimensional
isotropic  harmonic  oscillator  of  mass $\frac{1}{2}$  and  frequency
$B$ subjected  to an interaction involving  the (orbital) angular
momentum  $L$.  Defining the number operators $N_\pm= a_\pm^\dag a_\pm$ with
${a}_\pm=(a_y\pm ia_x)/\sqrt{2}$, we can write the  angular momentum and the Hamiltonian  operators as  $L=N_+ - N_-$
and $H_0= B \left( N_+ + N_- +1\right)$. As a result, these operators are diagonal
in  the  basis  of the number operators and lead to \cite{glr2,we3}
\be
\begin{array}{c}
  H_0 |n_+,n_->= B \left( n_+ + n_- +1 \right) |n_+,n_->\,,
  \\ \\
  L |n_+,n_-> = \left( n_+ -  n_- \right) |n_+,n_->\,,
  \\ \\
  {\cal H} |n_+,n_- > = B (2n_-+1) |n_+,n_->\,, \label{w3}
\end{array}
\ee
with $n_{\pm}=0,1,2,\cdots$.  We note  that the spectrum of ${\cal H}$
does  not   depend  on  the  label  $n_+$   and,  correspondingly,  its
eigenvectors are infinitely degenerate.

In the non-commutative space  the  Hamiltonian  for  the  SUSY  Landau  problem in the symmetric gauge
\eqref{symmgauge} is obtained from \eqref{h2-NC}  to be
\begin{equation}
2H^{\sc (NC)}=
    \left(\mathbf {p}^2    + \left(\frac{
    B}{2}\right)^2 \mathbf {x}^2 -
B L\right) \mathbf{1}_2 + {\cal F}_{12} \sigma_3 \,.\label{h2-NC1}
\end{equation}
We note here from \eqref{NCfieldstrength} that the symmetric gauge in \eqref{symmgauge} corresponds to a non-
commutative magnetic field of strength
\begin{equation}
{\cal F}_{12} = {\cal B} = B \left(1 + \frac{\theta B}{4}\right).\label{sub}
\end{equation}
Namely, the magnetic field in the non-commutative case is scaled by a factor of $\left(1 + \frac{\theta B}{4}\right)$ compared
to the commutative case. Furthermore, the Hamiltonian operator in \eqref{h2-NC1} acts on the wave function through the
Moyal product. Alternatively, we can use the Bopp shift \eqref{Bopp1}, to write an equivalent Hamiltonian which would act on
the wave functions through ordinary products. Note that under the Bopp shift \eqref{Bopp1}, we have
\begin{eqnarray}
\mathbf{p}^{2}\star \phi (\mathbf{x}) &=& \mathbf{p}^{2} \phi (\mathbf{x}),\nonumber\\
\mathbf{x}^{2} \star \phi (\mathbf{x}) &=& \left(\mathbf{x}^{2} + \frac{\theta^{2}}{4}\, \mathbf{p}^{2} - \theta L\right) \phi (\mathbf
{x}),\nonumber\\
L\star \phi (\mathbf{x}) &=& \left(L - \frac{\theta}{2}\, \mathbf{p}^{2}\right) \phi (\mathbf{x}),
\end{eqnarray}
so that the equivalent Hamiltonian with an ordinary product (see \eqref{h2-NC1}) can be written as
\begin{eqnarray}
2 H^{\sc (NC)} &=& \left(\mathbf{p}^{2} \left(1+ \frac{\theta B}{4}\right)^{2} + \left(\frac{B}{2}\right)^{2} \mathbf{x}^{2} - B \left(1+
\frac{\theta B}{4}\right) L\right) \mathbf{1}_{2} + {\cal B} \sigma_{3}\nonumber\\
& = &  \left(\overline{\mathbf{p}}^{2} + \left(\frac{\cal B}{2}\right)^{2} \overline{\mathbf{x}}^{2} - {\cal B} L\right) \mathbf{1}_{2} +
{\cal B}\,\sigma_{3},
\end{eqnarray}
where we have identified the scaled canonical variables
\begin{equation}
\overline{\mathbf{p}} = \mathbf{p}\left(1 + \frac{\theta B}{4}\right),\quad \overline{\mathbf{x}} = \frac{\mathbf{x}}{\left(1+\frac{
\theta B}{4}\right)}.\label{sub1}
\end{equation}
Thus,  in the variables \eqref{sub1}, the non-commutative SUSY  Landau problem  is equivalent  to the
standard SUSY Landau problem with  the magnetic field scaled as in
\eqref{sub}. The  energy spectrum and  the energy eigenvectors  can be
easily computed in terms of  those of an isotropic Harmonic oscillator
of   mass  $1/2$  and   frequency  $\omega=   {\mathcal  B}$,   as  in
(\ref{w3}). For the supersymmetric partner Hamiltonians we get
\begin{equation}
H^{\sc (NC)}_\pm \left| n_+ , n_-  ,\pm\right> = E_{n_-,n_+, \pm} \left| n_+
, n_- ,\pm\right> \,,
\end{equation}
with (see \eqref{w3})
\begin{equation}
    E_{n_-,n_+,\pm}  =  \frac{\cal B}{2} \left( 2 n_- +1\right) \pm \frac{\cal B}{2}\,,
\end{equation}
which does not depend on $n_+$, corresponding to an infinite degeneracy
of states.

\medskip

A second example is a particle moving in a linear superpotential
\begin{equation}
W_i = \frac{\alpha}{2}\, x_i,\label{linearpot}
\end{equation}
which  in  the  commutative  case leads  to  the SUSY isotropic two-dimensional
harmonic oscillator.  In this case, we take ${\rm Im}\, W_{i} = 0$ and
we identify ${\cal A}_{i} = -  \epsilon_{ij}\,{\rm Re}\, W^{j}$, so
that the supercharges can be written as
\begin{eqnarray}
Q    &    =     &    i    \left(p_{1}+ip_{2}    -    \frac{i\alpha}{2}
(x^{1}+ix^{2})\right)\sigma_{+},\nonumber
\\
Q^{\dagger}   &   =   &   -i\left(p_{1}-ip_{2}   +   \frac{i\alpha}{2}
(x^{1}-ix^{2})\right)\sigma_{-}.
\end{eqnarray}
Comparing these  supercharges with  \eqref{landauQ}, we see  that this
problem  can   be  mapped  to   the  SUSY  Landau  problem   with  the
identification  $\alpha =  B$  so  that the  earlier  analysis can  be
carried over in a straightforward manner.

\subsection{Three dimensional case}

To  construct a  general three  dimensional quantum  mechanical theory on a non-commutative space
which would provide a  realization of the algebra \eqref{H-SUSY-NC} we
follow the method  discussed in the previous section.  For example, we
introduce three arbitrary  superpotentials and define the supercharges
as  in  \eqref{susy3}  and   \eqref{SuperQ}.  The  nilpotency  of  the
supercharges
\begin{equation}
Q\star Q = (A_{k}\star  A_{l}) \psi_{k}\psi_{l} = 0 = Q^{\dagger}\star
Q^{\dagger}       =       (A^{\dagger}_{k}\star       A^{\dagger}_{l})
\psi^{\dagger}_{k}\psi_{l}^{\dagger},
\end{equation}
requires as before that
\begin{equation}
\psi_{k}\psi_{l} = 0 = \psi_{k}^{\dagger}\psi_{l}^{\dagger}.
\end{equation}
As a result,  \eqref{real} continues to provide a  realization for the
Grassmann  odd  elements  $\psi_{k},\psi^{\dagger}_{k},  k=1,2,3$.  As
before, we can consider a complex superpotential so that we can write
\begin{equation}
A_{k} = \partial_{k} + W_{k},\quad A_{k}^{\dagger} = -\partial_{k} + W_{k}^{*}.
\end{equation}

With  these the  Hamiltonian  for the  supersymmetric  theory, in  the
present case, can be obtained to be
\begin{eqnarray}
2 H^{\sc (NC)} & = & (A_{k}\star A_{l}^{\dagger}) \psi_{k}\psi^{\dagger}_{l}
+    (A_{l}^{\dagger}\star    A_{k})   \psi_{l}^{\dagger}\psi_{k}    =
\frac{1}{2}                           \{A_{k},A_{l}^{\dagger}\}_{\star}
\{\psi_{k},\psi_{l}^{\dagger}\}                                       +
\frac{1}{2}[A_{k},A_{l}^{\dagger}]_{\star}
     [\psi_{k},\psi^{\dagger}_{l}]\nonumber
\\
&=&  \left[  (\mathbf{p}  -  \mbox{\boldmath${\cal  A}$})_{\star}^2  +
  \left(     {\rm     Re}\,\mathbf{W}    \right)_\star^{2}     \right]
\mathbf{1}_2\otimes \mathbf{1}_{2} +  \left( \mathbf{D} \cdot {\rm Re}
\mathbf{W} \right)_{\star} \mathbf{1}_2\otimes \sigma_{3}
 +\epsilon_{klm}\left( -\frac{1}{2}   {\cal  F}_{kl} + i ({\rm Re} W_{k})\star ({\rm Re} W_{l})\right)\sigma_{m}\otimes \mathbf{1}_2
 \nonumber\\
&&  +
       \left[{\rm     Re}    \mathbf{W}    \times     (\mathbf{p}    -
         \mbox{\boldmath${\cal      A}$})     -      (\mathbf{p}     -
         \mbox{\boldmath${\cal   A}$})  \times  {\rm   Re}  \mathbf{W}
         \right]_{\star}    \cdot   \mbox{\boldmath$\sigma$}   \otimes
       \sigma_3\,,\label{NChamiltonian}
\end{eqnarray}
where we have identified
\begin{equation}
\mathbf{p}  = -  i\mbox{\boldmath$\nabla$},\quad \mbox{\boldmath${\cal
    A}$}    =   -   {\rm    Im}\,   \mathbf{W},\quad    \mathbf{D}   =
\mbox{\boldmath$\nabla$}  - i  \mbox{\boldmath${\cal  A}$},\quad {\cal
  F}_{kl} =  \partial_{k}{\cal A}_{l} - \partial_{l} {\cal  A}_{k} - i
\left[{\cal A}_{k},{\cal A}_{l}\right]_{\star}.
\end{equation}
For  any  arbitrary  complex  (three dimensional)  vector  superpotential
$\mathbf{W}   (\mathbf{x})$,   \eqref{NChamiltonian}  represents   the
Hamiltonian invariant under the supersymmetry transformation generated
by  the  Grassmann  odd   generators  $Q,  Q^{\dagger}$  (through  the
$\star$-operation).  Indeed, since the  Moyal product  is associative,
from (\ref{H-SUSY-NC}) we have
\begin{equation}\label{H-NC-inv-nilp}
    \left[  \varepsilon Q +  \overline{\varepsilon} Q^\dagger, H \right]_\star =
      \varepsilon \left[ \left(Q\right)_{\star}^{2}, Q^\dagger \right]_\star +
       \overline{\varepsilon}                                    \left[
         \left({Q^\dagger}\right)_{\star}^{2}, Q \right]_\star =0,
\end{equation}
for     arbitrary    Grassmann     parameters     $\varepsilon$    and
$\overline{\varepsilon}$  which  follows from  the  nilpotency of  the
supercharges guaranteed by the realization \eqref{psinilpotency}. (The
general  SUSY   transformations  in  the   non-commutative  space  are
discussed  in more detail  in appendix  \ref{NC-SUSY-Transform}). Let us note
that the  terms in  \eqref{NChamiltonian} can  be  rearranged to
identify  the  supersymmetric partner  Hamiltonians  in  the present case
to be
\begin{eqnarray}
2H_{\pm}^{\sc (NC)}   &  =   &  \left( (\mathbf{p}   -  \mbox{\boldmath${\cal
    A}$})^{2}_{\star}   +   ({\rm   Re}  \mathbf{W} )^{2}_{\star}   \pm
( \mathbf{D}\cdot       {\rm       Re}       \mathbf{W} )_{\star} \right)
\mathbf{1}_{2} +\epsilon_{klm}\left(- \frac{1}{2} {\cal F}_{kl} + i ({\rm Re} W_{k})\star ({\rm Re} W_{l})\right)\sigma_m
\nonumber\\
&&   \pm   \left(   {\rm   Re}   \mathbf{W}   \times   (\mathbf{p}   -
\mbox{\boldmath${\cal  A}$})  -  ( \mathbf{p}  -  \mbox{\boldmath${\cal
    A}$} )\times {\rm Re} \mathbf{W} \right)_{\star}\cdot{\boldsymbol \sigma}.
\label{susypair}
\end{eqnarray}

For completeness,  we note here that  the action of  the generators of supersymmetry on
functions of the position through  the $\star$-operation  can be obtained  by using the
Bopp shift defined in (\ref{Bopp}) as
\begin{eqnarray}
Q\star  \phi(\mathbf{x}) & = & A_k  \psi_k \star \phi(\mathbf{x})
      = (A_{k}\star  \phi (\mathbf{x})) \psi_{k} =  \Big( \partial_k +
      W_k\left( \mathbf{x} -
      \overline{\mathbf{p}}/2    \right)   \Big)\,   \phi(\mathbf{x})
      \sigma_k \otimes \sigma_+\,,\nonumber
\\
Q^\dagger  \star  \phi(\mathbf{x}) &  =  & A_k^\dagger  \psi_k^\dagger
\star\phi(\mathbf{x})
      = (A_{k}^{\dagger}\star \phi (\mathbf{x}) \psi_{k}^{\dagger}
      =  \Big(- \partial_k + W^{*}_k\left( \mathbf{x} -
      \overline{\mathbf{p}}/2    \right)   \Big)\,   \phi(\mathbf{x})
      \sigma_k \otimes \sigma_-\,\label{shifted}
\end{eqnarray}
and similarly for the $\star$-action of other operators on functions ($\overline{\mathbf{p}}$ is defined following \eqref{Bopp}).

With  this  construction  of  the  general  three  dimensional  SUSYQM
Hamiltonian  in the  non-commutative space,  we will  now  discuss one
simple example.  Although the  three-dimensional case in  principle is
straightforward, finding  soluble examples  is more difficult  than in
two-dimensions. We will  consider the
isotropic three-dimensional harmonic oscillator, for which the  superpotential is  given by
(see \eqref{linearpot})
\be
{\bf W}  ({\bf x}) =  \frac{\alpha}{2}\, {\bf x},
\ee
with $\alpha \in \mathbb{R}$. In this case, therefore, the superpotential is real and we have $\mbox{\boldmath${\cal
A}$} = - {\rm Im} \mathbf{W} = 0$. As a result, the Hamiltonian in
\eqref{NChamiltonian} takes the simpler form
\begin{eqnarray}
2H^{\sc (NC)} & = & \left(\mathbf{p}^{2} + ( \mathbf{W})^{2}_{\star}\right)
\mathbf{1}_{2}\otimes \mathbf{1}_{2} + (\mbox{\boldmath$\nabla$}\cdot
\mathbf{W}) \mathbf{1}_{2}\otimes \sigma_{3}\nonumber\\
& &  + i ( \mathbf{W} \times  \mathbf{W})_{\star} \cdot
\mbox{\boldmath$\sigma$} \otimes \mathbf{1}_{2} + ( \mathbf{W}\times
\mathbf{p} - \mathbf{p}\times  \mathbf{W})\cdot \mbox{\boldmath$\sigma$}
\otimes \sigma_{3}.\label{linearpot1}
\end{eqnarray}
The Moyal (star) products can be evaluated through the Bopp shift. In fact, noting
that in three dimensions, the parameter of non-commutativity can be identified with a
vector as
\begin{equation}
\theta^{ij} = \epsilon^{ijk} \theta_{k},
\end{equation}
under a Bopp shift we can write (see \eqref{Bopp})
\begin{equation}
f (\mathbf{x}) \rightarrow f (\mathbf{x} + \frac{1}{2}\,
\mbox{\boldmath$\theta$}\times \mathbf{p}).
\end{equation}
Using this the Hamiltonian in \eqref{linearpot1} can be written in the simple form
\begin{eqnarray}
2 H^{\sc (NC)} & = & \left(\left(1+ \left(\frac{\alpha}{4}\right)^{2}
\mbox{\boldmath$\theta$}^{2}\right) \mathbf{p}_{\perp}^{2} +
\mathbf{p}_{\parallel}^{2} + \frac{\alpha^{2}}{4}\,\mathbf{x}^{2} - \frac{\alpha^{2}}
{4}\, \mbox{\boldmath$\theta$}\cdot \mathbf{L}\right) \mathbf{1}_{2}\otimes
\mathbf{1}_{2} + \frac{3\alpha}{2}\, \mathbf{1}_{2}\otimes \sigma_{3}\nonumber\\
& & - \frac{\alpha^{2}}{4}\, \mbox{\boldmath$\theta$}\cdot \mbox{\boldmath$\sigma$}
\otimes \mathbf{1}_{2} + \frac{\alpha}{2}\left(2\mathbf{L} - \mathbf{p}^{2}\,
\mbox{\boldmath$\theta$} + (\mbox{\boldmath$\theta$}\cdot \mathbf{p})
\mathbf{p}\right)\cdot \mbox{\boldmath$\sigma$} \otimes \sigma_{3}.
\end{eqnarray}
Here we have denoted the parallel component of $\mathbf{p}$ along the direction $\mbox{\boldmath$\theta$}$ by
$\mathbf{p}_{\parallel} = \frac{\mbox{\boldmath$\theta$}\cdot \mathbf{p}}{\mbox{\boldmath$\theta$}^{2}}\,\mbox{\boldmath$
\theta$}$ while the perpendicular component is given by $\mathbf{p}_{\perp} = \mathbf{p} - \mathbf{p}_{\parallel}$ and the
orbital momentum is defined as usual as $\mathbf{L} = \mathbf{x}\times \mathbf{p}$. The supersymmetric partner
Hamiltonians (see \eqref
{susypair}) are now determined to be
\begin{eqnarray}
2H^{\sc (NC)}_{\pm} & = & \left(\left(1+ \left(\frac{\alpha}{4}\right)^{2} \mbox{\boldmath$\theta$}^{2}\right) \mathbf{p}_{\perp}^{
2} +
\mathbf{p}_{\parallel}^{2} + \frac{\alpha^{2}}{4}\,\mathbf{x}^{2} - \frac{\alpha^{2}}{4}\, \mbox{\boldmath$\theta$}\cdot
\mathbf{L} \pm \frac{3\alpha}{2}\right) \mathbf{1}_{2}\nonumber\\
& & +\left(-\frac{\alpha^{2}}{4}\, \mbox{\boldmath$\theta$} \pm  \frac{\alpha}{2}\left(2\mathbf{L} - \mathbf{p}^{2}\,\mbox{
\boldmath$\theta$} + (\mbox{\boldmath$\theta$}\cdot \mathbf{p}) \mathbf{p}\right)\right)\cdot \mbox{\boldmath$\sigma$}.
\end{eqnarray}
Even in this simple case, the energy spectrum and the eigenfunctions for the dynamical system cannot be obtained in closed
form in a
straightforward manner as in the two-dimensional case. A perturbative treatment, however, seems  possible.

\section{Conclusions}
In this paper, we have constructed systematically general supersymmetric quantum mechanical models in higher dimensional
non-commutative space. The general models require a restricted nilpotency condition for the Grassmann odd elements and
allow for complex superpotentials. This is consistent with the explicit realizations chosen in earlier studies where the nature of
the algebra was not fully appreciated. Non-commutativity of spatial coordinates is introduced through the (Moyal) $\star$-
product or through the Bopp shift. As we go to higher dimensions, we find that it is not easy to find solvable models. However,
all these models have the characteristic feature that they introduce interactions involving angular momentum. This can,
therefore, be considered as a starting point for several possible applications in diverse fields where the mixture
between   noncommutativity and supersymmetry could be of interest. For example nuclear, optics and atomic physics are
systems where many potentials involving spin and non-local interactions appear and the general framework developed in this
paper would possibly prove useful.
\vspace{0.4cm}

\noindent{\bf Acknowledgement}

\noindent This   work  was  partially  supported  by
FONDECYT-Chile and CONICYT under grants 1050114, 1060079, 7080062 and DICYT (USACH),
and by PIP 6160 - CONICET, and UNLP grant 11/X381, Argentina as well as by USDOE Grant number DE-FG 02-91
ER40685.

\appendix
\section{SUSY invariance and the nilpotency of $\psi_k$ and $\psi_k^\dagger$}\label{SUSYinvariance}
The aim of the present Appendix is to explicitly show that the nilpotency of $\psi_k$ and $\psi_k^\dagger$ is  sufficient to
ensure the SUSY invariance of the Hamiltonian, even though these elements do not necessarily satisfy a Clifford algebra (
see, for example, the explicit realization in \eqref{real}). For simplicity, we will describe the proof in the standard (commutative)
case with a real superpotential, the proof for the cmplex superpotential or the non-commutative case can be shown in a
parallel manner.

With the supercharges written as in (\ref{susy3}) and (\ref{SuperQ}) and assuming the nilpotency of the Grassmann odd
elements as in \eqref{psinilpotency} (without using the explicit realization \eqref{real}),
\begin{equation}\label{nil1}
    \psi_k\,\psi_l    =   0\,, \quad    \psi^{\dagger}_k\,\psi^{\dagger}_l =0 \,,
\end{equation}
we obtain the SUSY transformations for the basic elements to be
\begin{eqnarray}
\label{ts1}
\delta x_i &=& \left[\varepsilon  Q  +  \overline{\varepsilon}
      Q^\dagger,x_i \right] = \varepsilon\psi_i-\bar{\varepsilon}\psi^\dag_i \,,
\\
\delta    p_i  & =&\left[\varepsilon  Q  +  \overline{\varepsilon}
      Q^\dagger,p_i \right] =   i\,W_{j,i}    \big(\varepsilon\,\psi_j   +
\bar{\varepsilon }\,\psi^\dag_j\big)\,,
\\
\delta \psi_i &=&\left[\varepsilon  Q  +  \overline{\varepsilon}
      Q^\dagger,\psi_i \right] =  A^\dag_k \, \bar{\varepsilon} \left\{ \psi^\dag_k,\psi_i\right\}\,,
\\
\delta\psi_i^\dag &=&\left[\varepsilon  Q  +  \overline{\varepsilon}
      Q^\dagger,\psi_i^\dag \right]  = A_k\, {\varepsilon} \left\{\psi_k,\psi_i^\dag\right\}\,.
\end{eqnarray}
Here we have assumed that $Q$ generates supersymmetry transformations with the Grassmann parameter $\varepsilon$
while $Q^{\dagger}$ generates those with parameter $\overline{\varepsilon}$.

For simplicity, we put $\bar{\varepsilon}=0$ and consider only the change in the Hamiltonian in (\ref{H-SUSY}) coming from
transformations generated by $Q$ (proportional
to $\varepsilon$, the neglected terms can be obtained by taking the conjugates of the result obtained in the following),
\begin{eqnarray}
\label{varham}
4\delta H
&=&\bigg(\{(\partial_iW_k-\partial_kW_i),\partial_j-W_j\}+\{(\partial_jW_k     +     \partial_kW_j),\partial_i+W_i\}     \bigg)
{\varepsilon}\psi_k\{\psi_i,\psi_j^\dag\} \nonumber
\\
&&+\bigg([(\partial_iW_k-\partial_kW_i),\partial_j-W_j] -[(\partial_jW_k +\partial_k W_j),\partial_i+W_i]     \bigg)
{\varepsilon}\psi_k [\psi_i,\psi_j^\dag] \nonumber
\\
&&+\bigg(\{\partial_i+W_i,-\partial_j+W_j\} \{\psi_i,
[\varepsilon \psi_k,\psi_j^\dag]\} +  [\partial_i+W_i,-\partial_j+W_j] [\psi_i,
[\varepsilon\psi_k,\psi_j^\dag]]
\bigg)(\partial_k+W_k)\,.
\end{eqnarray}

In this expression the Grassmann odd  elements appear in combinations of products of the type
$\psi_k \psi_i \psi_j^\dag\,, \psi_j^\dag\psi_i \psi_k$ and $\psi_k\psi_j^\dag\psi_i$.
The  first  two types of terms vanish by virtue  of (\ref{nil1}), while the third  leads to
\begin{eqnarray}
2\delta H &=& \bigg(
(\partial_j-W_j)(\partial_iW_k-\partial_kW_i) + (\partial_jW_k +\partial_kW_j)(\partial_i+W_i) + \nonumber
\\
&&(-\partial_j+W_j)(\partial_i + W_i)(\partial_k+W_k)-(\partial_k+W_k)(-\partial_j+W_j)(\partial_i+W_i)
\bigg) {\varepsilon}\psi_ k\psi_j^\dag\psi_i \,. \nonumber
\end{eqnarray}
Noting  that
\begin{equation}
[\partial_k\mp W_k,\partial_j+W_j]= (\partial_kW_j\pm\partial_jW_k),
\end{equation}
we find
\begin{eqnarray}
2\delta       H      &=&\bigg([\partial_k-W_k,\partial_j+W_j]      +
[\partial_k+W_k,\partial_j-W_j]\bigg)(\partial_i+W_i)  {\varepsilon}\psi_ k\psi_j^\dag\psi_i=0\,.
\end{eqnarray}
Therefore, the nilpotency assumed in (\ref{nil1}) is a sufficient condition for the SUSY invariance of $H$.
Let us emphasize that  without   (\ref{nil1})  variations
proportional   to  the   products  of the type $\psi_k  \psi_i   \psi_j^\dag$ and $\psi_j^\dag\psi_i   \psi_k$   do not cancel by
themselves and a possibility for their cancellation would require the anticommutator  $\{\psi_i,\psi_j^\dag\}$ to be carefully
chosen in a nontrivial manner.

\section{SUSY transformation in non-commutative space}\label{NC-SUSY-Transform}

As discussed in appendix \ref{SUSYinvariance}, the nilpotency of the matrices $\psi_k$ and
$\psi_k^\dagger$ in (\ref{psinilpotency}) is a sufficient condition for the SUSY-invariance of the Hamiltonian in both, the
standard and the
non-commutative three-dimensional case.

\medskip

Let us now consider a general SUSY transformation in the three-dimensional non-commutative space, realized unitarily in
terms of the generators of transformation as
\begin{equation}
\label{SUSY-unitary}
    \left( e^{\varepsilon Q + \overline{\varepsilon} Q^\dagger} \right)_\star
    := 1 + \varepsilon Q + \overline{\varepsilon} Q^\dagger +
    \frac{1}{2} \left\{\varepsilon Q , \overline{\varepsilon} Q^\dagger \right\}_\star \,,
\end{equation}
where $\varepsilon$ and $\overline{\varepsilon}:= \varepsilon^\dagger$ are constant Grassmann odd parameters which
anticommute with
the supercharges. (The forms of the supercharges are described in sections {\bf II} and {\bf III}.) It can be checked easily that
\begin{equation}
\label{SUSY-unitary2}
      \left( e^{\varepsilon Q + \overline{\varepsilon} Q^\dagger} \right)_\star
    \star \left( e^{\varepsilon Q + \overline{\varepsilon} Q^\dagger} \right)_\star^\dagger = \left\{ 1 + \varepsilon Q + \overline{
    \varepsilon} Q^\dagger +
    \frac{1}{2} \left\{\varepsilon Q , \overline{\varepsilon} Q^\dagger \right\}_\star \right\}
    \star  \left\{ 1 -\overline{\varepsilon} Q^\dagger - \varepsilon Q +
    \frac{1}{2} \left\{\varepsilon Q , \overline{\varepsilon} Q^\dagger \right\}_\star \right\}
    =1  \,.
\end{equation}

The supersymmetric transformation of  functions  depending  on the dynamical variables   $x^{k},  \psi_k$ and
${\psi_k}^\dagger$  can be obtained from
\begin{equation}\label{t1}
      \left( e^{\varepsilon Q + \overline{\varepsilon} Q^\dagger} \right)_\star
\star F\left(x_k, p_k, \psi_k, {\psi_k}^\dagger \right) \star
\left( e^{\varepsilon Q + \overline{\varepsilon} Q^\dagger} \right)_\star^\dagger =  F +  \left[\varepsilon  Q + \overline{
\varepsilon} Q^\dagger, F\right]_\star
+ \frac{1}{2} \Big[\varepsilon  Q + \overline{\varepsilon} Q^\dagger,
\left[\varepsilon  Q + \overline{\varepsilon} Q^\dagger, F
  \right]_\star \Big]_\star  \simeq  F + \delta F\,,
\end{equation}
where we have identified the first order changes in the Grassmann parameters with
\begin{equation}\label{delta-F}
    \delta F =  \left[\varepsilon  Q + \overline{\varepsilon} Q^\dagger, F\right]_\star\, .
\end{equation}

\medskip

Explicitly, the SUSY transformations for the coordinates can be obtained from
\begin{equation}\label{t2}
      \delta  x_i  =   \left[\varepsilon  Q  +  \overline{\varepsilon}
      Q^\dagger,x_i \right]_\star = \left( \varepsilon \psi_i - \overline{\varepsilon} {\psi_i}^\dagger \right)
       - \imath \theta_{i j} \left(W_{k,j}(\mathbf{x})\, \varepsilon \psi_k +  W^{*}_{k,j} (\mathbf{x})\,\overline{\varepsilon} {\psi_k}^
       \dagger \right)\,,
\end{equation}
where we have denoted
\begin{equation}
W_{k,j} (x) = \partial_{j} W_{k} (\mathbf{x})\, .
\end{equation}
For the derivative operator we obtain
\begin{equation}\label{t3}
      \delta \partial_i = \left[\varepsilon Q + \overline{\varepsilon}
      Q^\dagger, \partial_i \right]_\star = -\left(W_{k,i}(\mathbf{x})\, \varepsilon \psi_k +  W^{*}_{k,i} (\mathbf{x}) \overline{
      \varepsilon} {\psi_k}^\dagger \right)\,.
\end{equation}
For the Grassmann odd elements in (\ref{su5}), relation \eqref{t1} yields for $\psi_i$
\begin{equation}\label{t4}
      \delta  \psi_i =  \left[\varepsilon  Q +  \overline{\varepsilon}
      Q^\dagger, \psi_i \right]_\star =  \left(-\partial_k + W_k^{*}(\mathbf{x})\right)  \overline{\varepsilon} \left\{   {\psi_k}^\dagger
      ,\psi_i \right\},
\end{equation}
while for $\psi_i^\dagger$ we have
\begin{equation}\label{t5}
      \delta     {\psi_i}^\dagger    =    \left[\varepsilon     Q    +
      \overline{\varepsilon} Q^\dagger, {\psi_i}^\dagger \right]_\star
      = \left(\partial_k + W_k(x)\right)  \varepsilon \left\{  \psi_k , {\psi_i}^\dagger \right\} \,,
\end{equation}
where the anticommutator $\left\{  \psi_k , {\psi_i}^\dagger \right\}$ is given in (\ref{conmut-psi}).
Let us note that in the limit $\theta_{ij}\rightarrow 0$, these expressions reduce to the transformations for the standard (
commutative) space. Finally, we note that since the $\star$-product is associative, we have
\begin{equation}\label{t6}
      \delta    (F    \star     G)    =    \left[\varepsilon    Q    +
      \overline{\varepsilon} Q^\dagger, F \star G \right]_\star = \delta F \star G + F \star \delta G\,.
\end{equation}


\end{document}